\documentclass[review]{elsarticle}

\usepackage{amssymb}
\usepackage{amsmath}

\usepackage{color}
\usepackage{graphicx}
\usepackage{bm}
\usepackage{color}

\journal{Physica E}

\begin{document}

\begin{frontmatter}



\title{Quantum Transport of Dirac fermions in graphene with a spatially varying Rashba spin-orbit coupling}


\author{Leila Razzaghi}
\author{Mir Vahid Hosseini\corref{cor1}}
\cortext[cor1]{Corresponding author.}
\ead{mv.hosseini@znu.ac.ir}

\address{Department of Physics, Faculty of Science, University of Zanjan, Zanjan 45371-38791, Iran}

\begin{abstract}
We theoretically study electronic transport through a region with inhomogeneous Rashba spin-orbit (RSO) coupling placed between two normal regions in a monolayer graphene. The inhomogeneous RSO region is characterized by linearly varying RSO strength within its borders and constant RSO strength in the central region. We calculate the transmission properties within the transfer matrix approach. It is shown that the amplitude of conductance oscillations reduces and at the same time the magnitude of conductance increases with increasing border thickness. We also investigate how the Fano factor can be modified by the border thickness of RSO region.
\end{abstract}


\end{frontmatter}


\section{Introduction}
\label{intro}
Quantum transport of graphene-based nanostructures has attracted much of interest due to potential applications \cite{QuanTrans}. One of the interesting features of graphene originates from the possibility of engineering the structural and electronic properties of graphene \cite{Graphene}. In this regard, although in pristine graphene, the strength of spin-orbit coupling is weak, but the possibility of enhancing the strength of spin-orbit coupling in graphene layer has recently attracted a great interest from theoretical \cite{SOtheor1,SOtheor2,SOtheor3,SOtheor4} and experimental \cite{SOexper1,SOexper2} viewpoints. It has been shown that spin-orbit coupling in graphene can be increased, giving rise to the so-called spin Hall effect \cite{SHE1,SHE2} at zero external magnetic fields at room temperature. This opens up new opportunity to manipulate charge- and spin-related phenomena efficiently same as manipulation of pseudospin \cite{Pseudospintronics} and valley \cite{valleytronics1,valleytronics2,valleytronics3,valleytronics4} degrees of freedom for carbon-based electronics applications \cite{GraphSpintronics}.

Basically, the spin-orbit coupling in graphene can be divided into intrinsic and extrinsic origins \cite{SHE2}. The intrinsic spin-orbit coupling comes from carbon intra-atomic spin-orbit coupling and it can be raised up to 17mev by proximity effect
between monolayer graphene and few-layers semiconducting tungsten disulfide \cite{proxiSO}. The extrinsic spin-orbit coupling, which is known as Rashba spin-orbit (RSO), can be induced through breaking the inversion symmetry of the lattice structure. A giant Rashba splitting has been observed as large as 100 $meV$ through Au intercalation at the graphene $-$ Ni interface \cite{giantSO}. In realistic cases, when RSO coupling induces in a specific region of a sample through external sources such as gate configuration \cite{RashbaGate1,RashbaGate2}, substrate \cite{RashbaSub1,RashbaSub2,RashbaSub3} or adatoms \cite{RashbaAdatoms1,RashbaAdatoms2,RashbaAdatoms3,RashbaAdatoms4}, the strength of RSO coupling may not change abruptly at the region's boundaries, and its strength modulates spatially from full value to the zero \cite{moduSO1,moduSO2}, due to the leakage of the source outside the region.

Transport properties under modulated potential \cite{pnPoten1,pnPoten2,pnPoten3} and energy gap \cite{gapvary1,gapvary2,gapvary3,gapvary4} in graphene layer have been considered, but the role of spatially varying RSO coupling on the transport properties of graphene still lacks a clear understanding \cite{nonuniform1,nonuniform2}.
While most of the studies of transport properties from RSO region focus on abrupt change of strength of spin-orbit coupling at the borders \cite{SOabrupt1,SOabrupt2,SOabrupt3,SOabrupt4,SOabrupt5,spin-chiral,termo}, in this paper, we will address the effects of variation of RSO coupling with a constant strength gradient within its border regions on the transmission probability, the conductance and the Fano factor of graphene. We find that the magnitude and the angular dependence of the transmission probabilities depend strongly on the border thickness. In addition, it is shown
that if the border thickness is finite, its effect on the conductance is not only to reduce the oscillations
amplitude but also to enhance the conductance magnitude. Also, depending on the values of border thickness, the Fano factor of the system can be controlled fundamentally.

\section{Model and Theory}
\label{sec:1}
\begin{figure}[t]
\begin{center}
\includegraphics[width=7.5cm]{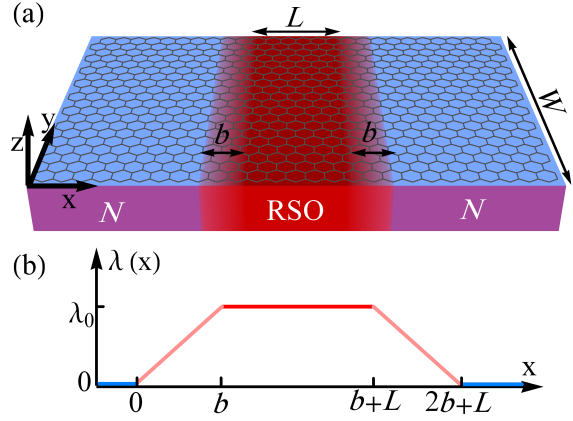}\\
\caption{(Color online) (a) Schematic representation of monolayer graphene with Rashba spin-orbit (RSO) region between two normal regions. The RSO region consists of the central region with constant RSO strength and border regions with smoothly varying strength of RSO coupling. The thickness of central and borders regions of RSO region are $L$ and $b$, respectively. Also, $W$ is the width of system. (b) Spatial profile of RSO region with linearly varying RSO strength in the border regions and constant strength $\lambda_0$ in the central region.
\label{SoSchem}}
\end{center}
\end{figure}

We consider a monolayer graphene in the $xy$ plane with an inhomogeneous RSO region between two normal regions. The interfaces of the regions are perpendicular to the $x$ direction as shown schematically in Fig. \ref{SoSchem}(a). The inhomogeneous RSO region is comprised of border regions that are close to the normal regions, characterizing by linearly varying strength of RSO coupling of thickness $b$ and the central region with thickness $L$ over which RSO strength is constant (see Fig. \ref{SoSchem}(b)). So the spatial profile of RSO strength throughout the spin-orbit region, 0 $\leq x \leq$ 2$b$ + $L$, can be modeled as,
\begin{eqnarray}\label{landa}
\lambda(x)=\Bigg\{
\begin{array}{cc}
  \lambda_0 \frac{x}{b}, & 0\leq x< b \\
  \lambda_0, & b\leq x\leq b+L \\
  \lambda_0 \frac{(L+2b-x)}{b}, & b+L< x\leq 2b+L
\end{array}
\end{eqnarray}
where $\lambda_0$ is the height of RSO strength. The spatial variation of RSO coupling is assumed to be small on the scale of the graphene's lattice constant (a = 0.246 $nm$). In this semiclassical approach, the long-wavelength components of Hamiltonian will be valid, subsequently, states near the valley K$+$ and those near K$-$ can be considered as two separate valleys in constructing the continuum approximation of Hamiltonian.
Also, due to the absence of inter-valley scattering processes, we adopt single valley picture.

Within the continuum limit, the Hamiltonian describing the total system reads as \cite{SOabrupt1,SOabrupt2,SOabrupt3,SOabrupt4,SOabrupt5,spin-chiral,termo}
\begin{eqnarray}\label{Hamiltonain}
H=&-&i\hbar v_f(\sigma_x\partial_x+\sigma_y\partial_y) s_0 \nonumber\\
  &+&\lambda(x)(\sigma_x s_y-\sigma_y s_x)\Theta(x)\Theta(L+2b-x),
\end{eqnarray}
where $v_f\approx10^6m/s$ is the Fermi velocity, the Pauli's matrices $\mathbf{\sigma}$ and $\mathbf{s}$ act on the pseudospin (sublattice) and spin spaces, respectively. Also, $\Theta(x)$ is the Heaviside step function and $s_0$ is the identity matrix.

Note that in conventional Rashba systems, Rashba Hamiltonian is momentum-dependent \cite{Winkler}, subsequently, in the case of non-uniform Rashba coupling, there will be terms in the Hamiltonian containing derivatives of the Rashba coupling in order to ensure hermiticity of the Rashba operator \cite{Winkler}. But in the case of graphene due to triangular symmetry, by expanding Rashba Hamiltonian around Dirac points (low-energy expansion), the lowest order of expansion is momentum-independent \cite{SHE2} which is in contrast to the conventional systems. However, if we retain higher order terms in the expansion which are momentum-dependent \cite{Rakyta}, the contribution of terms including derivatives of Rashba coupling is negligible, because in our analysis we have made the assumption that the strength of spin-orbit coupling changes linearly and smoothly.

For the numerical implementation, it is convenient to consider RSO region as a sequence of slices perpendicular to the direction of transport with nearly constant strength in each slice. By solving Eq. (\ref{Hamiltonain}) in the $i$th slice of RSO region, one can determine the eigenvalues as
\begin{eqnarray}\label{eigenenergies}
E^i_{l,n}=l\sqrt{\lambda^2_i+\hbar^2v^2_f(k^2+q^2)}+n\lambda_i,
\end{eqnarray}
where $n = + (-)$ represents upper (lower) subband in the $l = + (-)$ conduction (valance) band, $k$ ($q$) is the $x$ ($y$) component of wave vector which will be specified below and $\lambda_i$ is RSO strength in the $i$th slice. Also, the corresponding eigenvectors are given by,
\begin{eqnarray}\label{eigenvectors}
\psi^{i,m}_{l,n}\!&=&\!\frac{e^{imkx}e^{iqy}/2}{\sqrt{(E^i_{l,n})^2+\hbar^2v^2_f(k^2+q^2)}}\!
\left(\!\!
                                                                                         \begin{array}{c}
                                                                                           -in\hbar v_f(mk-iq) \\
                                                                                           E^i_{l,n} \\
                                                                                           -inE^i_{l,n} \\
                                                                                           \hbar v_f(mk-iq) \\
                                                                                         \end{array}
                                                                                       \!\!\right)
,\nonumber\\
\end{eqnarray}
where $m$ = 1 (-1) indicates the right-moving (left-moving) carrier.
In the left and right normal regions, Eqs. (\ref{eigenenergies}) and (\ref{eigenvectors}) reduce, respectively, as
\begin{eqnarray}\label{Neigenenergies}
E_{l}=l\hbar v_f\sqrt{k^2+q^2},
\end{eqnarray}
\begin{eqnarray}\label{Neigenvectors}
\psi^{N,m}_{l,n}(x,y)&=&\frac{e^{imkx}e^{iqy}}{2}
\left(
                                                                                         \begin{array}{c}
                                                                                           -inme^{im\phi} \\
                                                                                           l \\
                                                                                           -iln \\
                                                                                           m e^{im\phi} \\
                                                                                         \end{array}
                                                                                       \right)
,
\end{eqnarray}
where $\phi=\tan^{-1}q/k$.

In the calculation, an incident carrier is supposed to be injected from the left to the right upon incidence
angle $\phi$. Due to particle-hole symmetry in the system, without loss of generality, we focus on the conduction band states, $l = +$, and choose a positive value for the carrier energy ($E > 0$). Translational invariance in the $y$ direction implies that the $y$ component of the wave vector is a good quantum number, which it can be written in terms of the incidence angle $\phi$ as
\begin{equation}\label{eq:q}
q=\frac{E}{\hbar v_f} \sin\phi\,.
\end{equation}
But the $x$ component of wave vector depends on the region under consideration. Using Eqs. (\ref{eigenenergies}), (\ref{Neigenenergies}) and (\ref{eq:q}), the $x$ component of wave vector can be obtained as

\begin{equation}\label{eq:kN}
k=E\sqrt{(1-\sin^2\phi)}/\hbar v_f\,,
\end{equation}
for the normal regions, and
\begin{equation}\label{eq:k}
k^{in}=\sqrt{E(E-2 n\lambda_i-E \sin^2\phi)}/\hbar v_f\,
\end{equation}
for the $i$th slice of the RSO region. From Eq. (\ref{eq:k}), it is easy to see that for $n = -$ and $E > 0$, $k^{in}$ has pure real values, consequently, we have traveling modes. But for $n = +$ and $0 < E < 2 \lambda^i$, $k^{in}$ becomes pure imaginary and the wave is evanescent, whereas for $n = +$ and $E > 2 \lambda^i$ the wave is a traveling mode\cite{termo,evan}.

Upon substituting Eqs. (\ref{eq:kN}) and (\ref{eq:q}) into Eq. (\ref{Neigenvectors}), one can use Eq. (\ref{Neigenvectors}) to write the general wave functions which are valid in the left and right normal regions as,
\begin{eqnarray}
\Psi(x \leq 0)&=&\psi_{+n}^{N+}+ r_{n,-}\psi_{+-}^{N-}+r_{n,+}\psi_{++}^{N-}\\
\Psi(L+2b\leq x)&=&t_{n,-}\psi_{+-}^{N+}+t_{n,+}\psi_{++}^{N+},
\end{eqnarray}
where the coefficient $t_{n,m}$ ($r_{n,m}$) represents the transmission (reflection) amplitude from incident subband $n$ into subband $m$.
Also, with the help of Eq. (\ref{eq:k}), (\ref{eq:q}) and (\ref{eigenvectors}), the total wave function in the $i$th slice of RSO region can be expressed as,
\begin{eqnarray}\label{eq:f2}
\Psi(0\leq x \leq L+2b)&=&a^i_{n,-}\psi_{+-}^{i+}+a^i_{n,+}\psi_{++}^{i+}\nonumber\\
&+&b^i_{n,-}\psi_{+-}^{i-}+b^i_{n,+}\psi_{++}^{i-}\,,
\end{eqnarray}
where the coefficients $a^i_{n,m}$ and $b^i_{n,m}$ are the scattering amplitudes from subband $n$ into subband $m$.

Using the transfer-matrix method \cite{transfer-matrix}, transmission amplitude in the left normal region, $t_{n,m}$, can be determined. Then transmission probability $T_{n,m}$ can be evaluated by $T_{n,m}$ = $\mid t_{n,m}\mid^2$.
It should be noted that since the RSO interaction does not couple the states of spin-chiral carriers with opposite subband indices, so transmission probabilities through different subbands vanish, i.e., $T_{+,-}$ = 0 and $T_{-,+}$ = 0 \cite{spin-chiral}.
Having obtained transmission probabilities, the conductance of carriers at zero temperature in subband n can be obtained by \cite{landauer},
\begin{equation}\label{eq:G}
G_n=G_0 \int^{\pi/2}_{-\pi/2}  T_{n,n}(E_F,\phi) \cos\phi\,d\phi\,,
\end{equation}
where $G_0=2 e^2 W E_F/h^2 v_f$, $W$ is the width of system, $e$ is the electron charge, and  $E_F$ is the Fermi energy.

Finally, we investigate the subband Fano factor of this system which can be given as follows \cite{Fano},
\begin{equation}\label{eq:G}
F_n=\frac{\int^{\pi/2}_{-\pi/2}T_{n,n}(E_F,\phi)(1-T_{n,n}(E_F,\phi)) \cos\phi\,d\phi}{\int^{\pi/2}_{-\pi/2}T_{n,n}(E_F,\phi) \cos\phi\,d\phi}\,.
\end{equation}

\section{Numerical results and discussions}\label{sec3}
Throughout this paper, the value of $\lambda_0$ is fixed to be 10 $meV$ and $L$ = 100 $nm$. In the numerical calculations, we have chosen the thickness of each slice 1$nm$. It should be noted that we have examined larger slice thickness and same qualitative results were found.

\begin{figure}[t]
\begin{center}
\includegraphics[width=6.5cm]{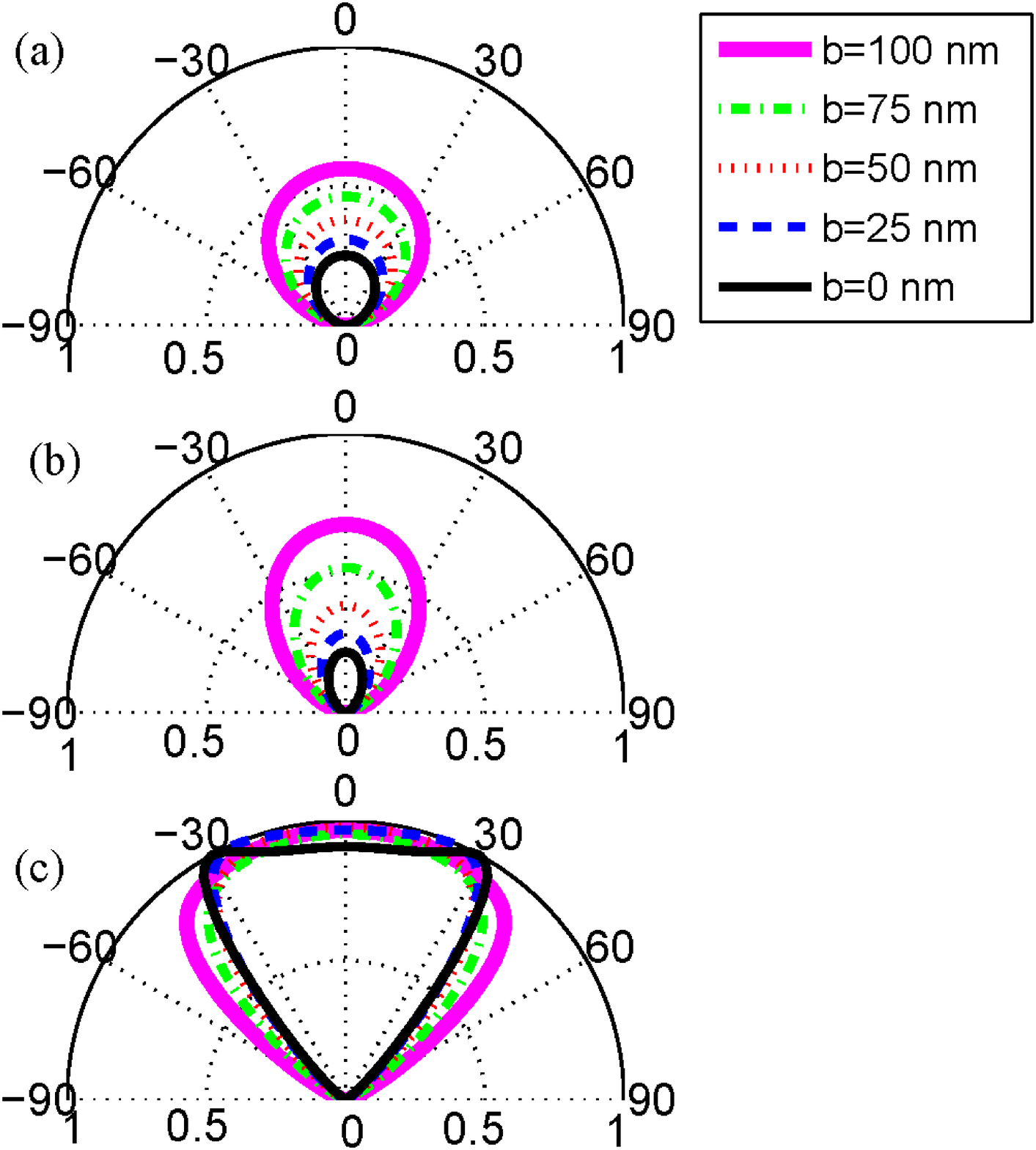}
\caption{(Color online) Angular dependence of transmission probability $T_{+,+}$ for different values of border thickness, $b$ = 0, 25, 50, 75, and 100 $nm$ at incident energy (a) $E$ = 2 $meV$, (b) $E$ = 16 $meV$ and (c) $E$ = 40 $meV$. Here $\lambda_0$ = 10 $meV$ and $L$ = 100 $nm$.
\label{Tuu}}
\end{center}
\end{figure}

Figures \ref{Tuu}(a), (b) and (c) represent the angular dependence of the transmission probability $T_{+,+}$ through RSO region for different values of border thickness, $b$ = 0, 25, 50, 75, and 100 $nm$. The incident energy in panel (a) is $E$ = 2 $meV$ and in panel (b) is $E$ = 16 $meV$. In both of these panels, we can see that the magnitude of $T_{+,+}$ increases with increasing the border thickness with a maximum at $\phi$ = 0 but at large value of $b$ the maximum value of $T_{+,+}$ in panel (b) is larger than that of panel (a). Panel (c) refers to the case $E$ = 40 $meV$ and illustrates that the angles at which transmission maxima occur depend on the border thickness. For large $b$ the transmission is close to unity at low angles, while for small $b$ perfect transmission takes place at $\phi\simeq \pm30^{\circ}$.

\begin{figure}[t]
\begin{center}
\includegraphics[width=6.5cm]{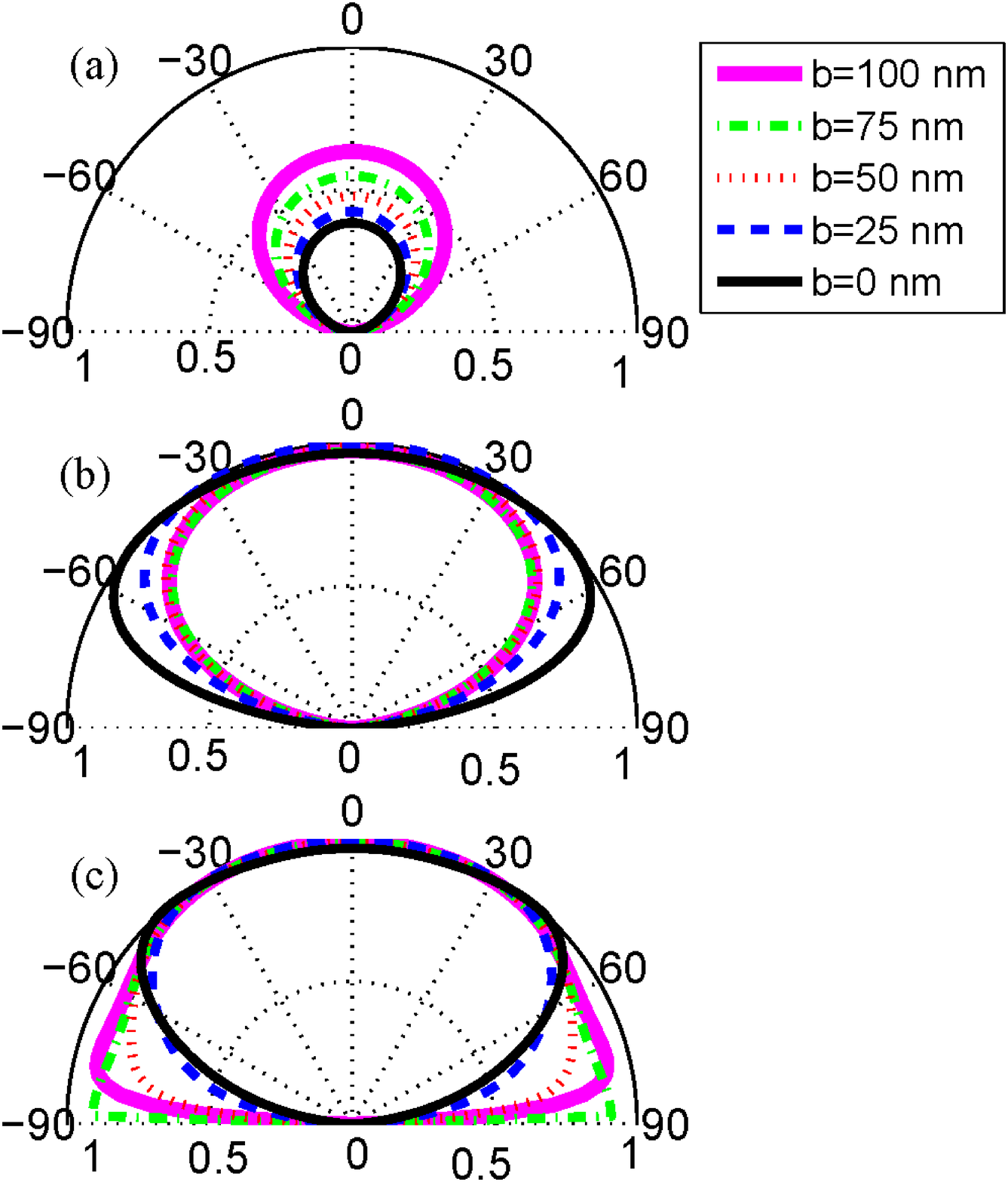}
\caption{(Color online) Angular dependence of transmission probability $T_{-,-}$ for different values of border thickness, $b$ = 0,25, 50, 75, and 100 $nm$ at incident energy (a) $E$ = 2 $meV$, (b) $E$ = 16 $meV$ and (c) $E$ = 40 $meV$. Here $\lambda_0$ = 10 $meV$ and $L$ = 100 $nm$.
\label{Tdd}}
\end{center}
\end{figure}

In Fig. \ref{Tdd} the transmission probability $T_{-,-}$ as a function of the incidence angle with the same parameters used for Fig. \ref{Tuu} is shown. Figure \ref{Tdd}(a) displays an increasing behavior of the $T_{-,-}$ in terms of $b$ which is similar to what is also observed for the $T_{+,+}$ at low energies (see also Fig. \ref{Tuu}(a)). But for energies $E$ = 16 $meV$ and $E$ = 40 $meV$, as illustrated in Fig. \ref{Tdd}(b) and (c), the $T_{-,-}$ is close to unity in a considerable range of incidence angles about the $\phi = 0^{\circ}$. Note that in Fig. \ref{Tdd}(b) the transmission angles of the $T_{-,-}$ decrease slightly by increasing the $b$. Hence the different behaviors observed for the transmission probabilities with respect to the border thickness, depend on the value of the incident energy. These features will have direct influences on the conductance properties which are investigated below.

\begin{figure}[t]
\begin{center}
\includegraphics[width=7.5cm]{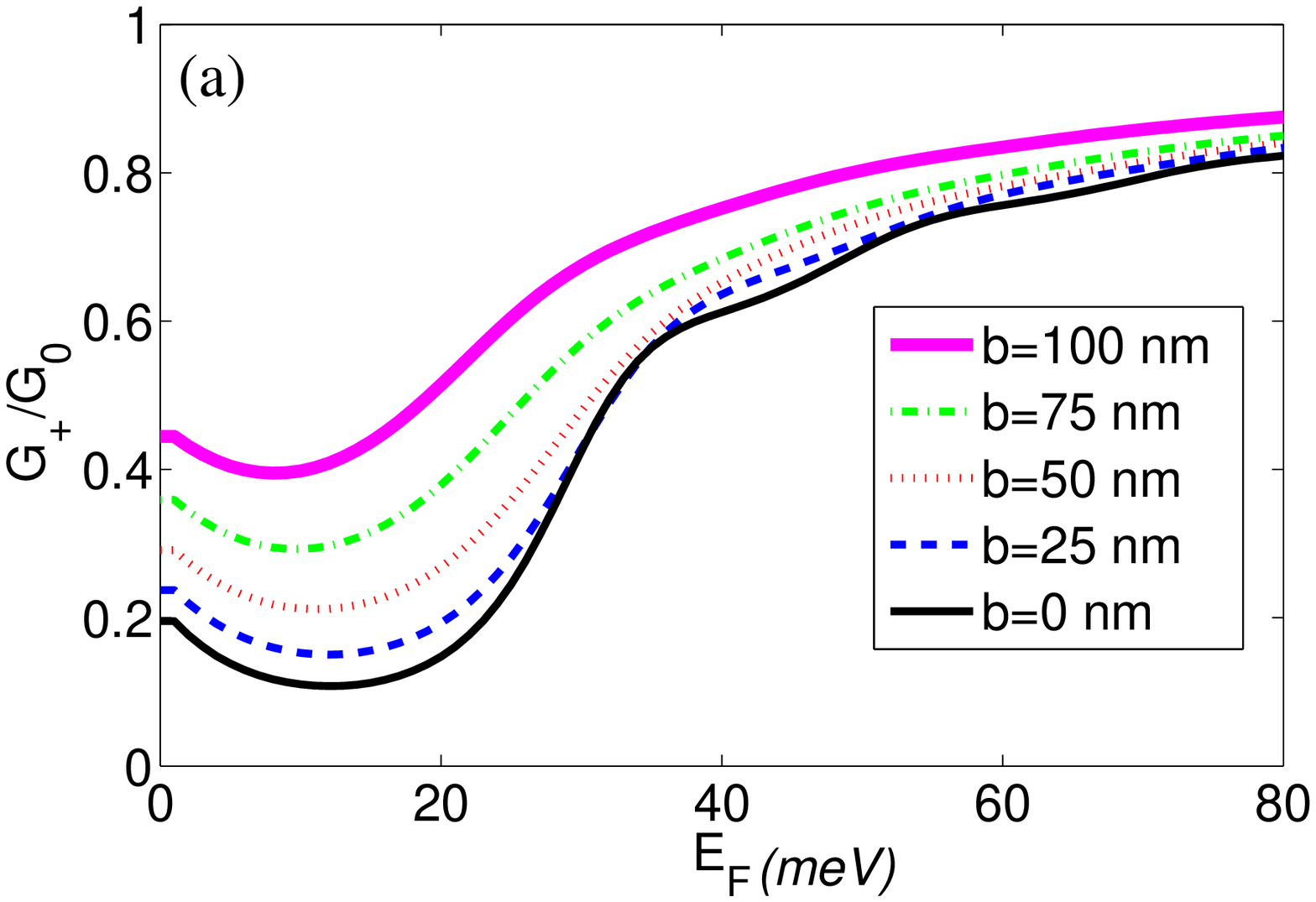}\\
\includegraphics[width=7.5cm]{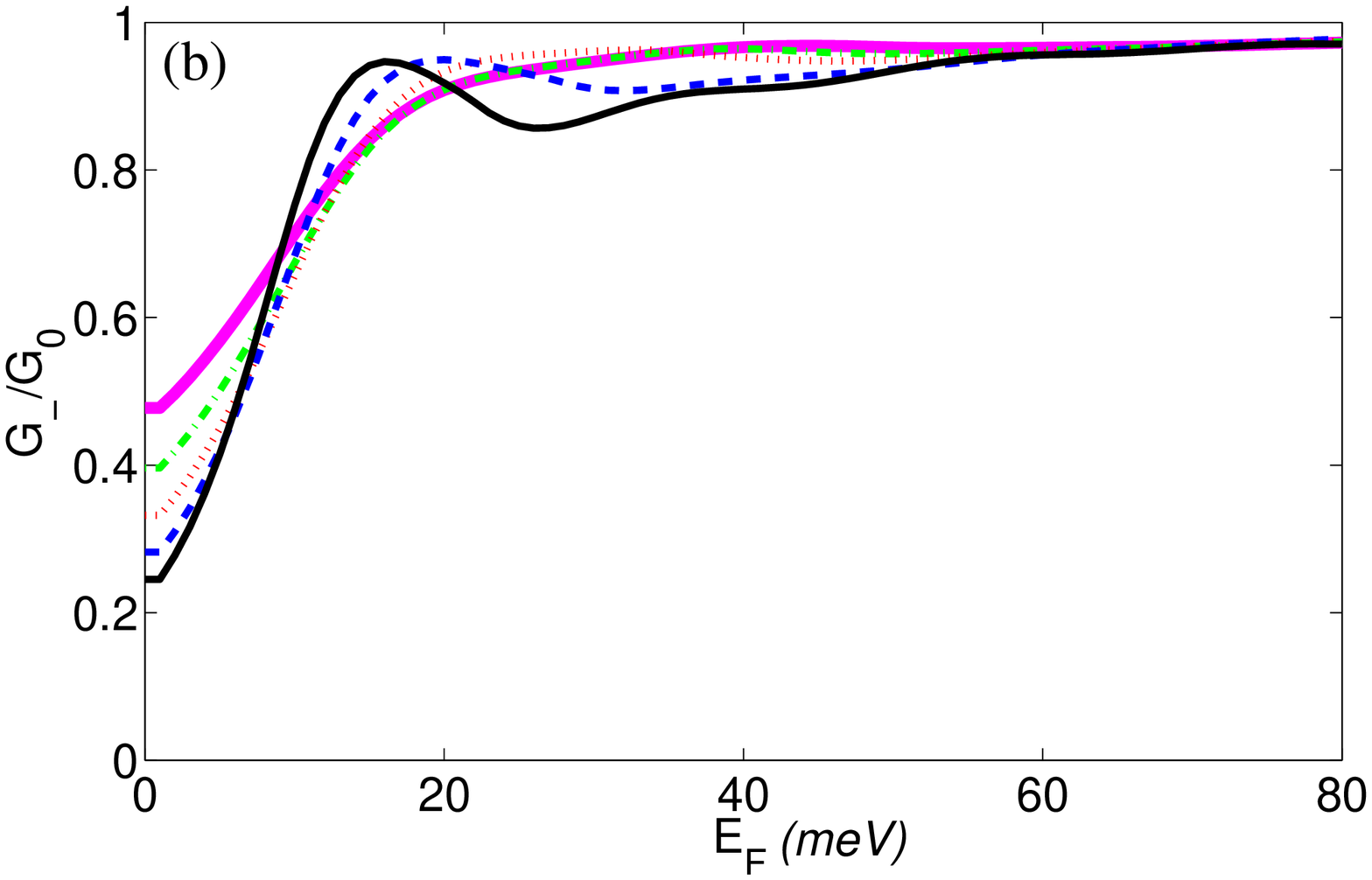}
\caption{(Color online) Conductance through (a) upper subband, $G_+$, and (b) lower subband, $G_-$  as a function of Fermi
energy for different values of border thickness, $b$ = 0, 25, 50, 75, and 100 $nm$. Here $\lambda_0$ = 10 $meV$ and $L$ = 100 $nm$.
\label{Gud}}
\end{center}
\end{figure}

In Fig. \ref{Gud}, the conductance of conduction subbands is presented as a function of the Fermi
energy for different values of the border thickness, $b$ = 0, 25, 50, 75, and 100 $nm$. The conductance of the upper subband, $G_+$, is shown in Fig. \ref{Gud} (a). We can see that the $G_+$ increases with increasing the border thickness, $b$. Nevertheless, this increase has not same value across the range of energies. For energies between 0 and $2\lambda_0$ in which carrier transmission can be occurred through evanescent modes, there exists a considerable enhancement in $G_+$ versus $b$. This indicates that, in contrast to the abrupt border case ($b$ = 0), the role of evanescent modes on transport features is less dominant when the border of RSO region is a smoothly varying function. In addition, for $E_F\gg 2\lambda_0$, a small enhancement can be observed in the $G_+$ corresponding to the large $b$, since, in this energy range scattering from spin-orbit interaction has no fundamental role.
Figure \ref{Gud}(b) shows the conductance of lower conduction subband, $G_-$. For $b$ = 0, $G_-$ increases and tends to the $G_0$ in a damped oscillatory behavior as a function of the Fermi energy. At low energies, increasing the border thickness causes an increase in the $G_-$. For intermediate Fermi energies, the modulations of $G_-$ reduce with an increasing the $b$ which this behavior is described below. Also, the values of $G_-$ are insensitive to the border thickness and take the maximal value $G_0$ in the limit $E_F\gg \lambda_0$. As a result, the $G_-$ increases monotonically as a function of the Fermi energy at sufficiently large $b$. Note that due to transmission of carrier via a traveling wave in the lower subband, the $G_-$ is often close to the $G_0$.

\begin{figure}[t]
\begin{center}
\includegraphics[width=7.5cm]{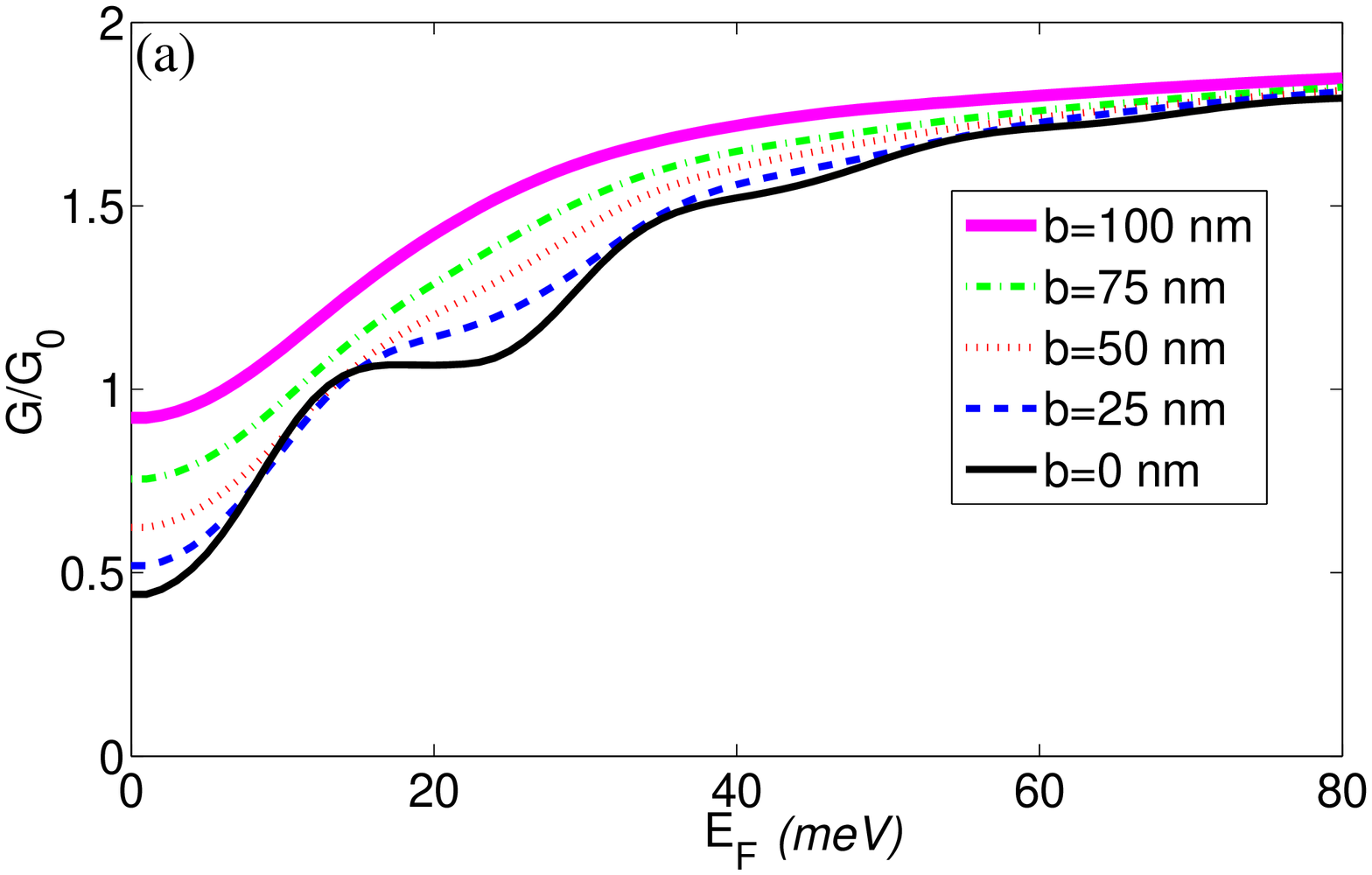}
\includegraphics[width=7.5cm]{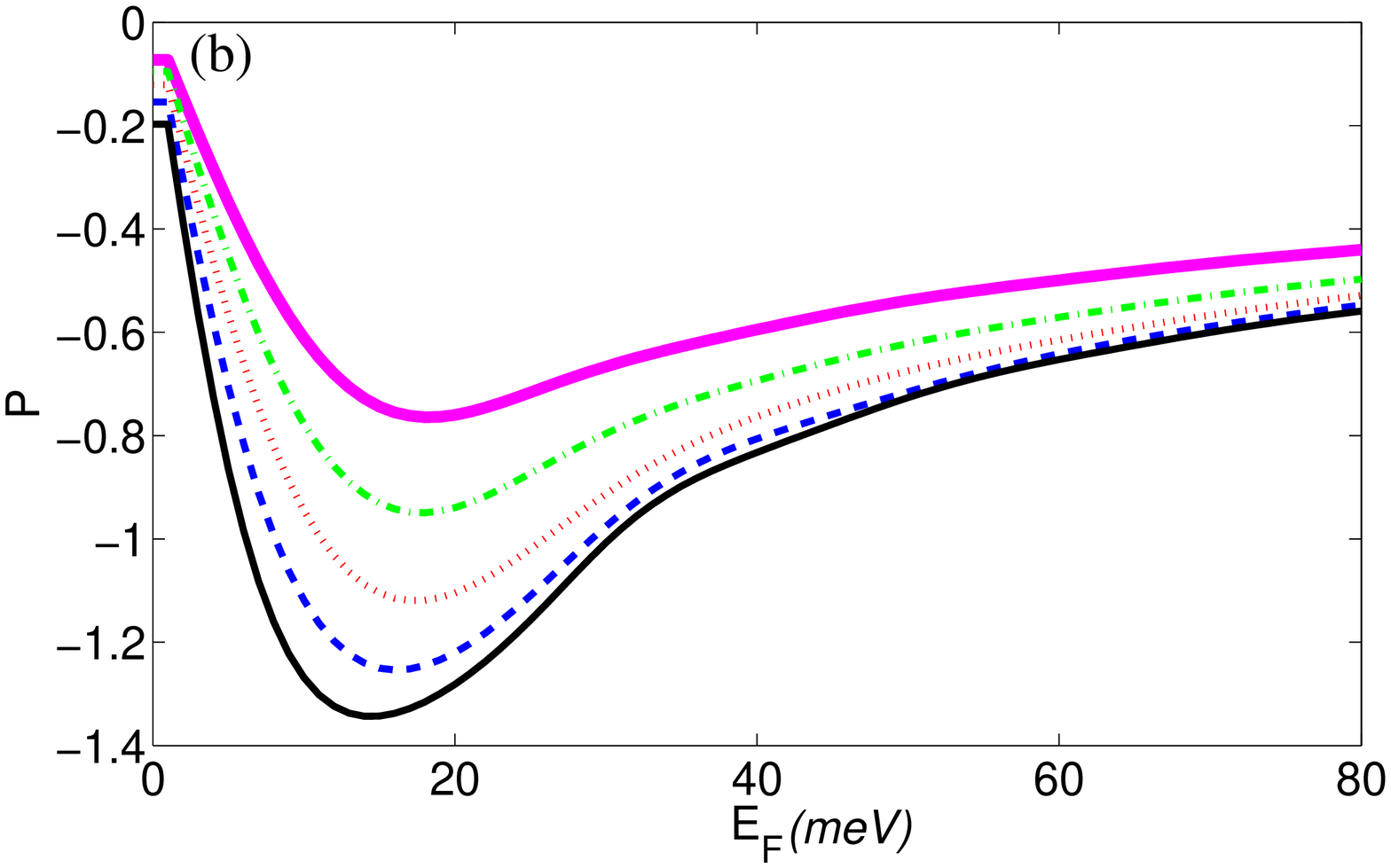}
\caption{(Color online) (a) Total conductance, $G = G_+ + G_-$  and (b) subband polarization of conductance $p = (G_+ - G_-)/G$ versus Fermi
energy for different values of border thickness, $b$ = 0, 25, 50, 75, and 100 $nm$. Here $\lambda_0$ = 10 $meV$ and $L$ = 100 $nm$.
\label{GTP}}
\end{center}
\end{figure}

The total conductance which is the sum of the conductances of the subbands, $G = G_+ + G_-$, is shown in Fig. \ref{GTP}(a). It increases versus Fermi energy because there always exist more traveling states at high energies. In addition, there are conductance oscillations which can be related to the interference between left-moving and right-moving states in the RSO region. With increasing $b$, $G$ not only increases but also displays damped oscillations. Remarkably, for large enough values of $b$ the oscillating behavior of conductance greatly reduces. This happens because when passing the carriers through the borders, they do not feel abrupt changes in dispersion relation due to smoothly varying spin-orbit coupling. Therefore reflection from the interfaces between the central and the normal regions decreases in comparison with abrupt border case and consequently the modulations of amplitude reduce.

Also, in Fig. \ref{GTP}(b), the subband polarization of the conductance $p = (G_+ - G_-)/(G_+ + G_-)$ versus Fermi energy is presented. We can see that there is a minima in the range $\lambda_0<E_F<2\lambda_0$. This indicates that in this energy range most of the carriers can be transmitted through the subband $n=-$. As we increase $b$ the polarization decreases in absolute value. It should be note that for $b$ = 0 all of the results presented in Figs. \ref{Gud} and \ref{GTP} (shown with thin solid black line) are consistent with Figs. 4 and 5 of Ref. \cite{termo}.

\begin{figure}[t]
\begin{center}
\includegraphics[width=7.5cm]{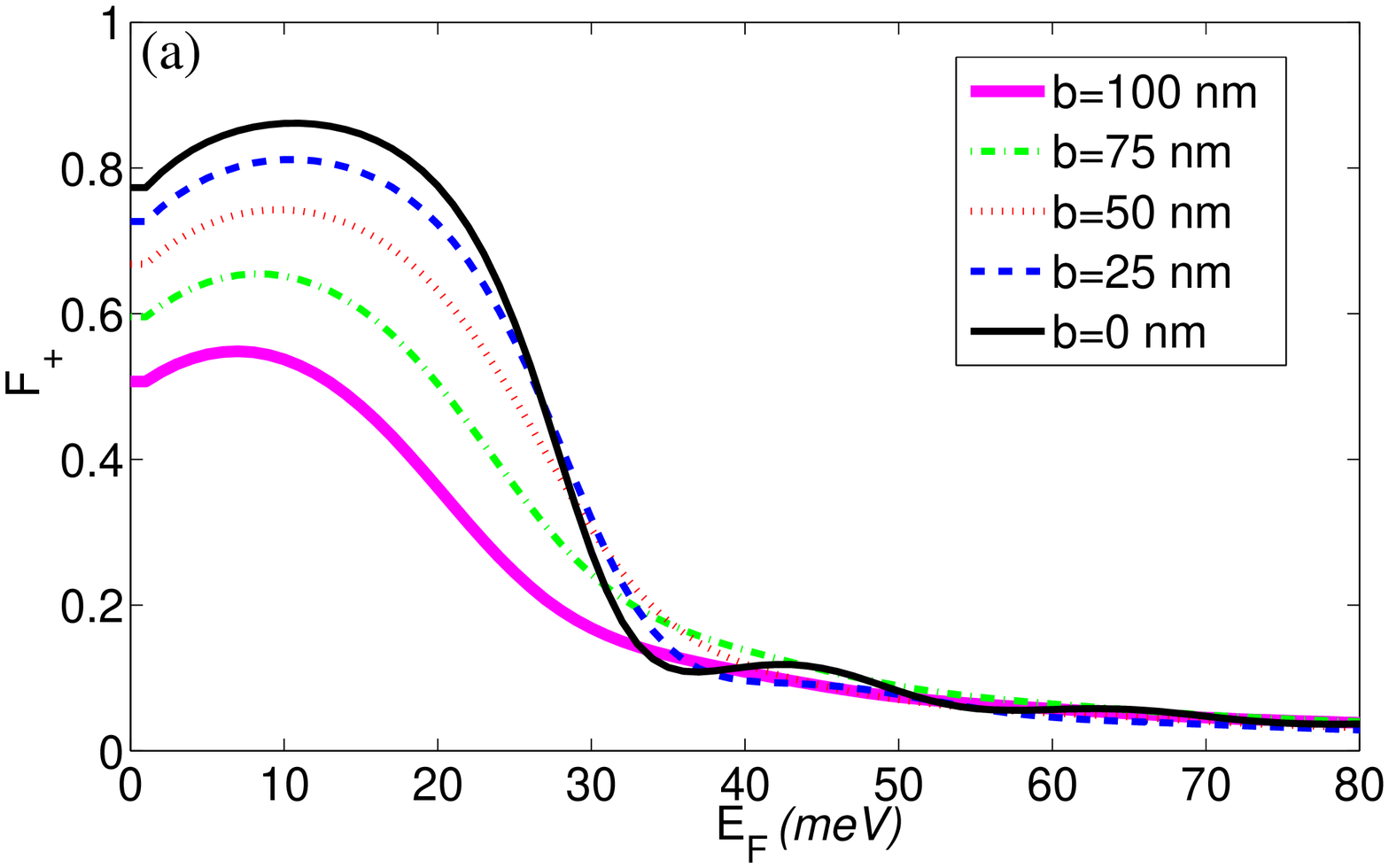}
\includegraphics[width=7.5cm]{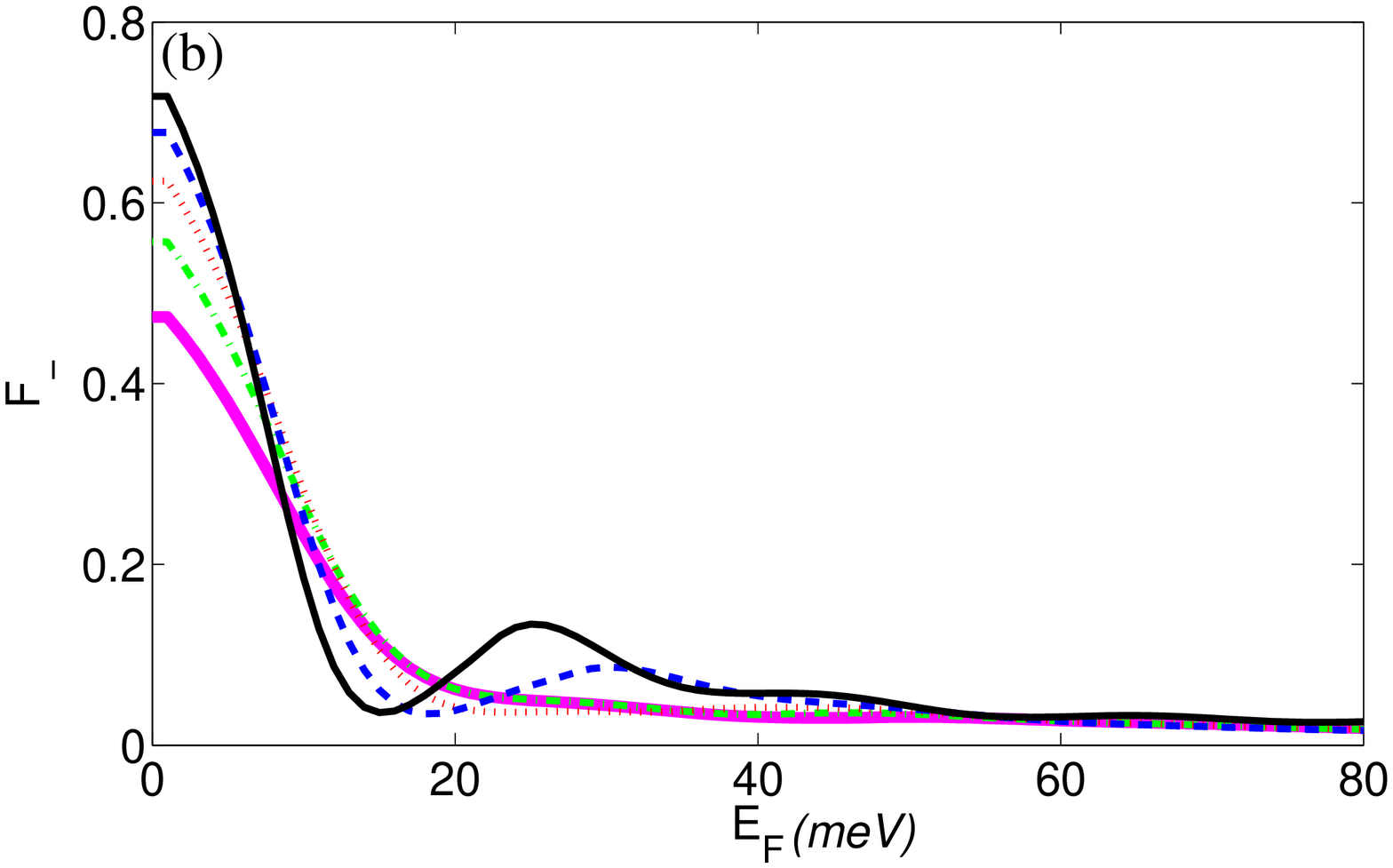}
\caption{(Color online) Fano factors of (a) upper subband, $F_+$  and (b) lower subband $F_-$ as a function of Fermi
energy for different values of border thickness, $b$ = 0, 25, 50, 75, and 100 $nm$. Here $\lambda_0$ = 10 $meV$ and $L$ = 100 $nm$.
\label{Fano}}
\end{center}
\end{figure}

\begin{figure}[t]
\begin{center}
\includegraphics[width=7.5cm]{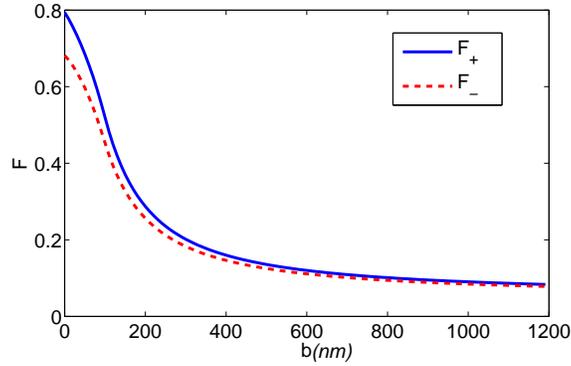}
\caption{(Color online) Fano factors of upper subband $F_+$ and lower subband $F_-$ as a function of border thickness $b$ at Fermi energy $E_F$ = 2 $meV$. Here $\lambda_0$ = 10 $meV$ and $L$ = 100 $nm$.
\label{Fanob}}
\end{center}
\end{figure}

Let us now investigate the Fano factor of the system providing insight into the internal kinetic rates. For the sake of clarity, the Fano factors of upper subband, $F_+$, and lower subband, $F_-$, are calculated separately which are shown in Figs. \ref{Fano}(a) and (b), respectively. At low energies, we observe that in both cases the effect of border thickness is to decrease the values of the Fano factors in particular for the $F_+$. Similar to the conductance behavior [Figs. \ref{GTP}(a) and (b) ], for intermediate energies, damped oscillatory behaviors of $F_{\pm}$ are reduced by increasing $b$. Also, for $E_F \gg \lambda_0$, transport characteristic turns into the ballistic regime ($F_{\pm}$ = 0).
Since at low energies spin-orbit scattering plays a dominant role, we focus on investigating the evolution of both the Fano factors ($F_{\pm}$) versus $b$ near the Dirac point. In Fig. \ref{Fanob}, this evolution is shown for $E_F$ = 2 $meV$.
We see that both the Fano factors are monotonically
decreasing function of the Fermi energy and tend to $F_{\pm}$ $\lesssim$ 0.1 for large enough values of $b$.
This implies that, in addition to the Fermi energy, the border thickness can influence on transport features in a fundamental way.

\section{Conclusion}\label{sec4}
In conclusion, we studied transport properties through inhomogeneous RSO region in graphene. We investigated transmission probabilities in terms of border thickness, Fermi energy and incidence angle. Also, in the analysis of the conductance as a function of the Fermi energy, we identified that the main influence of border thickness of RSO region is the smearing out of the conductance oscillations and at the same time the making an enhancement of conductance magnitude. Also, the effect of border thickness is to reduce the magnitude and oscillations amplitude of both Fano factors of upper and lower subbands. Further, we analyzed the low energy behaviors of the Fano factors for different subbands in a wide range of border thicknesses ranging from small thickness at which transmission takes place via tunneling effect, to a large one at which transport is nearly ballistic.

\section{Acknowledgements}
The authors are grateful to A. Darudi for fruitful discussions.


\begin{thebibliography}{50}

 \bibitem{QuanTrans}S. D. Sarma, S. Adam, E. H. Hwang, and E. Rossi, Rev. Mod. Phys. {\bf 83} (2011) 407.
 \bibitem{Graphene}A. H. Castro Neto, F. Guinea, N. M. R. Peres, K. S. Novoselov, and A. K. Geim, Rev. Mod. Phys. {\bf 81} (2009) 109.
\bibitem{SOtheor1}D. Huertas-Hernando, F. Guinea, and A. Brataas, Phys. Rev. B {\bf 74} (2006) 155426.
\bibitem{SOtheor2}J.-S. Jeong, J. Shin, and H.-W. Lee, Phys. Rev. B {\bf 84} (2011) 195457.
\bibitem{SOtheor3}W. Conan, H. Jun, A. Jason, F. Marcel, and W. Ruqian, Phys. Rev. X {\bf 1} (2011) 021001.
\bibitem{SOtheor4}K.-H. Jin, and S.-H. Jhi, Phys. Rev. B {\bf 87} (2013) 075442.

\bibitem{SOexper1}J. Balakrishnan, G. K. W. Koon, M. Jaiswal, A. H. Castro Neto, and B. \"{O}zyilmaz, Nat. Phys. {\bf 9} (2013) 284.
\bibitem{SOexper2}J. Balakrishnan, G. K. W. Koon, A. Avsar, Y. Ho, J. H. Lee, M. Jaiswal, S.-J. Baeck, J.-H. Ahn, A. Ferreira, M. A. Cazalilla, A. H. Castro Neto, and B. \"{O}zyilmaz, Nat. Commun. {\bf 5} (2014) 4748.


\bibitem{SHE1}J. E. Hirsch, Phys. Rev. Lett. {\bf 83} (1999) 1834.
\bibitem{SHE2}C. L. Kane, and E. J. Mele, Phys. Rev. Lett. {\bf 95} (2005) 226801.

\bibitem{Pseudospintronics}P. San-Jose, E. Prada, E. McCann, and H. Schomerus, Phys. Rev. Lett. {\bf 102} (2009) 247204.

\bibitem{valleytronics1}A. Rycerz, J. Tworzyd{\l}o, and C. W. J. Beenakker, Nat. Phys. {\bf 3} (2007) 172.
\bibitem{valleytronics2}Z. Z. Zhang, K. Chang, and K. S. Chan, Appl. Phys. Lett. {\bf 93} (2008) 062106.
\bibitem{valleytronics3}J. L. Garcia-Pomar, A. Cortijo, and M. Nieto-Vesperinas, Phys. Rev. Lett. {\bf 100} (2008) 236801.
\bibitem{valleytronics4}J. M. Pereira Jr., F. M. Peeters, R. N. Costa Filho, and G. A. Farias, J. Phys.: Condens. Matter {\bf 21} (2009) 045301.

\bibitem{GraphSpintronics}W. Han, R. K. Kawakami, M. Gmitra, and J. Fabian, Nat. Nanotechnol. {\bf 9} (2014) 794.

\bibitem{proxiSO}A. Avsar, J. Y. Tan, J. Balakrishnan, G. K. W. Koon, J. Lahiri, A. Carvalho, A. S. Rodin, T. Taychatanapat, E. C. T. O’Farrell, G. Eda, A. H. Castro Neto, and B. \"{O}zyilmaz, Nat. Commun. {\bf 5} (2014) 4875.

\bibitem{giantSO}D. Marchenko, A. Varykhalov, M. R. Scholz, G. Bihlmayer, E. I. Rashba, A. Rybkin, A. M. Shikin, and O. Rader, Nat. Commun. {\bf 3} (2012) 1232.

\bibitem{RashbaGate1}E.I. Rashba, Phys. Rev. B {\bf 68} (2003) 241315.
\bibitem{RashbaGate2}E.I. Rashba, Sov. Phys. Solid State {\bf2} (1960) 1109.

\bibitem{RashbaSub1}Yu. S. Dedkov, M. Fonin, U. R\"{u}diger, and C. Laubschat, Phys. Rev. Lett. {\bf 100} (2008) 107602.
\bibitem{RashbaSub2}A. Varykhalov, D. Marchenko, M. R. Scholz, E. D. L. Rienks, T. K. Kim, G. Bihlmayer, J. S\'{a}nchez-Barriga, and O. Rader, Phys. Rev. Lett. {\bf 108} (2012) 066804.
\bibitem{RashbaSub3}O. Rader, A. Varykhalov, and J. S\'{a}nchez-Barriga, Phys. Rev. Lett. {\bf 102} (2009) 057602.

\bibitem{RashbaAdatoms1}A. H. Castro Neto, and F. Guinea, Phys. Rev. Lett. {\bf 103} (2009) 026804.
\bibitem{RashbaAdatoms2}D. Ma, Zh. Li, and Zh. Yang, Carbon {\bf 50} (2012) 297.
\bibitem{RashbaAdatoms3}M. Gmitra, D. Kochan, and J. Fabian, Phys. Rev. Lett. {\bf 110} (2013) 246602.
\bibitem{RashbaAdatoms4}D. V. Fedorov, M. Gradhand, S. Ostanin, I. V. Maznichenko, A. Ernst, J. Fabian, and I. Mertig, Phys. Rev. Lett. {\bf 110} (2013) 156602.

\bibitem{moduSO1}S. J. Gong, and Z. Q. Yang. J. Phys. Condens. Matter {\bf 19} (2007) 446209.
\bibitem{moduSO2}A. Brataas, A. G. Mal'shukov, and Y. Tserkovnyak, New J. Phys. {\bf 9} (2007) 345.

\bibitem{pnPoten1}V. V. Cheianov, and V. I. Fal’ko, Phys. Rev. B {\bf 74} (2006) 041403.
\bibitem{pnPoten2}J. Cayssol, B. Huard, and D. Goldhaber-Gordon, Phys. Rev. B {\bf 79} (2009) 075428.
\bibitem{pnPoten3}M. -H. Liu, J. Bundesmann, and K. Richter, Phys. Rev. B {\bf 85} (2012) 085406.

\bibitem{gapvary1}L. Vitali, C. Riedl, R. Ohmann, I. Brihuega, U. Starke, and K. Kern, Surf. Sci. {\bf 602} (2008) L127.
\bibitem{gapvary2}G. Giavaras, and F. Nori, Appl. Phys. Lett. {\bf 97} (2010) 243106.
\bibitem{gapvary3}G. Giavaras, and F. Nori, Phys. Rev. B {\bf 83} (2011) 165427.
\bibitem{gapvary4}F. Zhai, and K. Chang, Phys. Rev. B {\bf 85} (2012) 155415.

\bibitem{nonuniform1}M. Rataj, and J. Barna\'{s}, Appl. Phys. Lett. {\bf 99} (2011) 162107.
\bibitem{nonuniform2}K. Hasanirokh, A. Phirouznia, F. Hassanirokh, and H. Mohammadpour, Appl. Phys. A {\bf117} (2014) 1963.

\bibitem{SOabrupt1}A. Yamakage, K.-I. Imura, J. Cayssol, and Y. Kuramoto, EPL {\bf 87} (2009) 47005.
\bibitem{SOabrupt2}D. Bercioux, and A. De Martino, Phys. Rev. B {\bf 81} (2010) 165410.
\bibitem{SOabrupt3}Ch. Bai, J. Wangb, Sh. Jia, and Y. Yang, Physica E {\bf 43} (2011) 884.
\bibitem{SOabrupt4}M. Esmaeilzadeh, and S. Ahmadi, J. Appl. Phys. {\bf 112} (2012) 104319.
\bibitem{SOabrupt5}Q. Zhang, K. S. Chanb, Z. Lin, and J.-F. Liu, Phys. Lett. A {\bf 377} (2013) 632.

\bibitem{spin-chiral}Kh. Shakouri, M. Ramezani Masir, A. Jellal, E. B. Choubabi, and F. M. Peeters, Phys. Rev. B {\bf 88} (2013) 115408.

\bibitem{termo}M. I. Alomar, and David S\'{a}nchez, Phys. Rev. B {\bf 89} (2014) 115422.

\bibitem{Winkler}R. Winkler, Spin Orbit coupling effects in two dimensional electron and hole systems, Springer, Berlin 2003.

\bibitem{Rakyta}P. Rakyta, A. Korm\'{a}nyos, and J. Cserti, Phys. Rev. B {\bf 82} (2010) 113405.

\bibitem{evan}M. Khodas, A. Shekhter, and A. M. Finkelstein, Phys. Rev. Lett. {\bf 92} (2004) 086602.

\bibitem{transfer-matrix}P. A. Mello, and N. Kumar, Quantum Transport in Mesoscopic Systems, Oxford University Press, Oxford 2004.

\bibitem{landauer}M. B\"{u}ttiker, Y. Imry, R. Landauer, and S. Pinhas, Phys. Rev. B {\bf 31} (1985) 6207.

\bibitem{Fano} J. Tworzyd{\l}o, B. Trauzettel, M. Titov, A. Rycerz, and C. W. J. Beenakker, Phys. Rev. Lett. {\bf 96} (2006) 246802.



\end{thebibliography}



\newpage

\end{document}